\documentstyle[a4wide,epsf]{article}

\newcommand{\lsim}{\mbox{\raisebox{-.9ex}{~$\stackrel{\mbox{$<$}}{\sim}$~}}}
\newcommand{\gsim}{\mbox{\raisebox{-.9ex}{~$\stackrel{\mbox{$>$}}{\sim}$~}}}
 
\begin{document}

\begin{center}
{\Large\bf Hybrid Inflation without Flat Directions and\\ 
without Primordial Black Holes}

\bigskip

{\Large  
Konstantinos Dimopoulos$^{*\dag}$
and Minos Axenides$^*$ 
}

\bigskip

$^*${\it 
Institute of Nuclear Physics, National Center for Scientific Research
`Demokritos',\\ 
Agia Paraskevi Attikis, Athens 153 10, Greece}\\
$^\dag${\it
Physics Department, Lancaster University, Bailrigg, Lancaster LA1 4YB, U.K.}

\begin{abstract}
We investigate the possibility that the Universe may inflate due to 
moduli fields, corresponding to flat directions of supersymmetry, lifted by 
supergravity corrections. 
Using a hybrid-type potential we obtain a two-stage inflationary model.
Depending on the curvature of the potential the first stage corresponds to
a period of fast-roll inflation or a period of `locked' inflation, 
induced by an oscillating inflaton. This is followed by a second stage of 
fast-roll inflation. We demonstrate that these two consecutive inflationary
phases result in enough total e-foldings to encompass the cosmological scales. 
Using natural values for the parameters (masses of order TeV and vacuum energy 
of the intermediate scale corresponding to gravity mediated supersymmetry 
breaking) we conclude that the $\eta$-problem of inflation is easily overcome. 
The greatest obstacle to our scenario is the possibility of copious production
of cosmologically disastrous primordial black holes due to the phase transition
switching from the first into the second stage of inflation. We study 
this problem in detail and show analytically that there is ample parameter 
space where these black holes do not form at all.
To generate structure in the Universe we assume the presence of a curvaton 
field. Finally we also discuss the moduli problem and how it affects our 
considerations.
\end{abstract}

\end{center}

\section{Introduction}

The latest elaborate observations of the anisotropy in the Cosmic Microwave 
Background Radiation (CMBR) suggest that structure formation in the Universe
is due to the existence of a {\em superhorizon} spectrum of curvature/density
perturbations, which are predominantly adiabatic and Gaussian \cite{wmap}.
The best mechanism to explain such perturbations is through the 
amplification of the quantum fluctuations of a light scalar field during a 
period of inflation. Moreover, inflation is to date the only compelling 
mechanism to account for the horizon and the flatness problems of the 
Standard Hot Big Bang (SHBB) cosmology (for a review see 
\cite{book0}\cite{book1}\cite{book}).

However, despite its successes, inflation remains as yet a paradigm without a 
model. According to this paradigm, inflation is realised through the 
domination of the Universe 
by the potential density of a light scalar field, which is slowly rolling down
its almost flat potential. One of the reasons for using a flat potential is 
that one requires inflation to last long enough for the cosmological scales
to exit the horizon during the period of accelerated expansion, so as to solve 
the horizon and flatness problems. This requires the inflationary period to 
last at least 40 e-foldings while, in most models, this number is increased to 
60 or more \cite{book}. Hence, the potential density of the
field has to remain approximately constant (to provide the effective 
cosmological constant responsible for the accelerated expansion) for some time,
which renders inflation with steep potentials unlikely. Thus, inflation seems
to require the presence of a suitable flat direction in field space. 

Unfortunately, flat directions are very hard to attain in supergravity because 
K\"{a}hler corrections generically lift the flatness of the scalar potential.
This is the so-called $\eta$-problem of inflation \cite{randall}. 
To overcome this problem many authors assume that accidental cancellations 
minimise the K\"{a}hler corrections or consider directions whose flatness is 
protected by a symmetry other than supersymmetry, such as, for example, the 
Heisenberg symmetry in no-scale supergravity models \cite{noscale} or a global 
U(1) for Pseudo-Nambu-Goldstone Bosons (PNGBs) (natural inflation 
\cite{natural}). However, accidental cancellations or no-scale supergravity 
require special forms for the K\"{a}hler potential, which have, to date, 
little theoretical justification. Also, inflation due to a PNGB suffers from 
other problems; for example in the limit of unbroken U(1) symmetry the PNGB 
potential vanishes. Consequently, there seems to be a generic problem in 
realising inflation without tuning.

Still, there have been attempts to overcome this problem.
A first step toward inflation without a flat direction was achieved by the 
so-called fast-roll inflation, introduced in Ref.~\cite{fastroll}. There, it 
has been shown that, even if the curvature of the inflaton potential is 
comparable to the Hubble parameter, one may have inflation for a limited 
number of e-foldings. However, this number turns out to be rather small and 
appears to reduce drastically if the effective mass of the inflaton increases. 
Hence, fast-roll inflation alone is probably not capable to explain the 
observations. It is possible, however, that it may assist other types of 
inflation, which are also incapable to last long enough. A prominent
example is thermal inflation \cite{thermal}, which also does not use a flat 
direction but suffers from the disadvantage of requiring the presence of a 
thermal bath preexisting inflation, to which the inflaton is strongly coupled
\cite{liber}. 

Recently, however, a new mechanism for inflation without a flat direction
was suggested in Ref.~\cite{dvali}. According to this mechanism, inflation 
may be achieved in a hybrid-type potential (introduced originally in the 
slow-roll hybrid inflation model \cite{hybrid}), where the field's rapid 
oscillations keep the former onto an unstable saddle point and prevent it from 
rolling toward the true minima. As pointed out in Ref.~\cite{dvali}, such type 
of potentials are natural for moduli fields. Unfortunately, the number of 
e-foldings of oscillatory inflation was found to be insufficient to solve the 
horizon and flatness problems. Due to this fact the authors of 
Ref.~\cite{dvali} introduced a metastable local minimum in the potential, 
rendering their model a two-stage inflation. During the first stage of 
inflation, the field sits in the metastable minimum and the Universe undergoes 
a period of so-called old inflation \cite{old}. This period terminates when 
the field tunnels out into the second stage of oscillatory inflation. 

In this paper we suggest a simpler and more generic scenario for inflation
without a flat direction. We point out that, in a hybrid-type non-flat 
potential one can have two consecutive stages of fast-roll inflation. However, 
if the curvature of the first stage of inflation is larger than a critical 
value, then this stage turns into a period of oscillatory `locked' inflation, 
which provides a lower bound on the number of e-foldings corresponding to this
inflationary stage. After the first stage of inflation, it is possible to 
have a second period of fast-roll inflation between the time when the field 
leaves the unstable saddle point until it rolls to the true minimum. This 
second stage may last long enough to enable the total inflationary period to 
solve the horizon and flatness problems, without imposing stringent bounds on 
the curvature of the potential. 

The biggest obstacle for our scenario to work is the possibility of copious
production of Primordial Black Holes (PBHs) due to the phase transition that 
switches from the first stage of inflation to the second one. This danger has 
been already identified in \cite{bastards}. Fortunately, we have found a way 
to circumvent the problem and avoid PBH production altogether. 

Since our inflaton is not a light field it cannot be responsible for the
generation of the observed superhorizon spectrum of curvature perturbations.
We, therefore, consider that these curvature 
perturbations are due to a curvaton field \cite{curvaton}, which has no
contact with the inflaton sector and cannot affect in any way the 
inflationary dynamics. We show that it is possible to achieve the necessary
e-foldings of inflation using natural values for the model parameters.

Our paper is structured as follows: In Sec.~2 we present a simple version
of non-flat, modular hybrid inflation. We show that this is a two stage
inflationary model. Depending on the curvature of the potential, the first 
stage can be a period of either fast-roll or oscillatory, `locked' inflation. 
We study in detail both cases and obtain an estimate of the number 
of e-foldings using natural values for the model parameters. 
In Sec.~3 we focus on the second stage of inflation, which corresponds to 
tachyonic fast-roll inflation, that uses the waterfall field of hybrid 
inflation as an inflaton. We find the number of e-foldings corresponding to 
this second inflationary phase and, hence, we obtain the total number 
of e-foldings of inflation. 
In Sec.~4 we compare the total number of e-foldings from both
the stages of inflation to the e-foldings corresponding to the largest
cosmological scales. Thereby, we calculate the bound on the 
tachyonic mass of the inflaton, which ensures enough e-foldings of inflation
to solve the horizon and flatness problems. In Sec.~5 we present a detailed 
analysis of the disastrous possibility of PBH production and offer a natural 
solution, which allows ample parameter space. In particular, we study carefully
the evolution and initial conditions of both moduli during the first stage of 
inflation and show that, if their interaction is not too strong, it is quite 
possible to avoid altogether the generation of PBHs. In Sec.~6 we discuss other
cosmological aspects of our models such as the generation of density 
perturbations using a curvaton field or the moduli problem. Finally, in
Sec.~7 we discuss our results and present our conclusions.

Throughout our paper we use natural units such that \mbox{$\hbar=c=1$} and
Newton's gravitational constant is \mbox{$8\pi G=m_P^{-2}$}, where 
\mbox{$m_P=2.4\times 10^{18}$GeV} is the reduced Planck mass.

\section{Fast--Roll versus Locked Inflation}

Consider two
moduli fields, which parameterise supersymmetric flat directions (whose 
flatness is lifted by supergravity corrections) with a
hybrid type of potential of the form
\begin{equation}
V(\Phi,\phi)=\frac{1}{2}m_\Phi^2\Phi^2+
\frac{1}{2}\lambda\Phi^2\phi^2+\frac{1}{4}\alpha(\phi^2-M^2)^2,
\label{V}
\end{equation}
where $\Phi$ and $\phi$ above are taken to be real scalar fields and
\begin{eqnarray}
m_\Phi\sim\frac{M_S^2}{m_P}\sim m_{3/2}\;, &
\qquad M\sim m_P\;,\qquad &
\alpha\sim\left(\frac{M_S}{m_P}\right)^4,
\label{param}
\end{eqnarray}
with \mbox{$\lambda\leq 1$} and
\mbox{$M_S\sim\sqrt{m_{3/2}m_P}\sim 10^{5.5}$GeV} being the intermediate 
scale corresponding to gravity mediated supersymmetry breaking, where
\mbox{$m_{3/2}\sim$ TeV} stands for the electroweak scale (gravitino mass). 
From the above we see that the tachyonic mass of $\phi$ is given by
\begin{equation}
m_\phi\sim\sqrt{\alpha}\,M\sim m_{3/2}
\label{mphi0}
\end{equation}
and its self-coupling is 
suppressed gravitationally as expected, \mbox{$\alpha\sim(m_{3/2}/m_P)^2$}. 

The above potential has global minima at \mbox{$(\Phi,\phi)=(0,\pm M)$} and
an unstable saddle point at \mbox{$(\Phi,\phi)=(0,0)$} similarly to hybrid 
inflation (see Figure~1). 
However, in contrast to regular hybrid inflation, for 
\mbox{$|\Phi|,|\phi|\leq m_P$} the potential does not satisfy the slow-roll
requirements. 

\begin{figure}
\begin{center}

\leavevmode
\hbox{%
\epsfxsize=4in
\epsffile{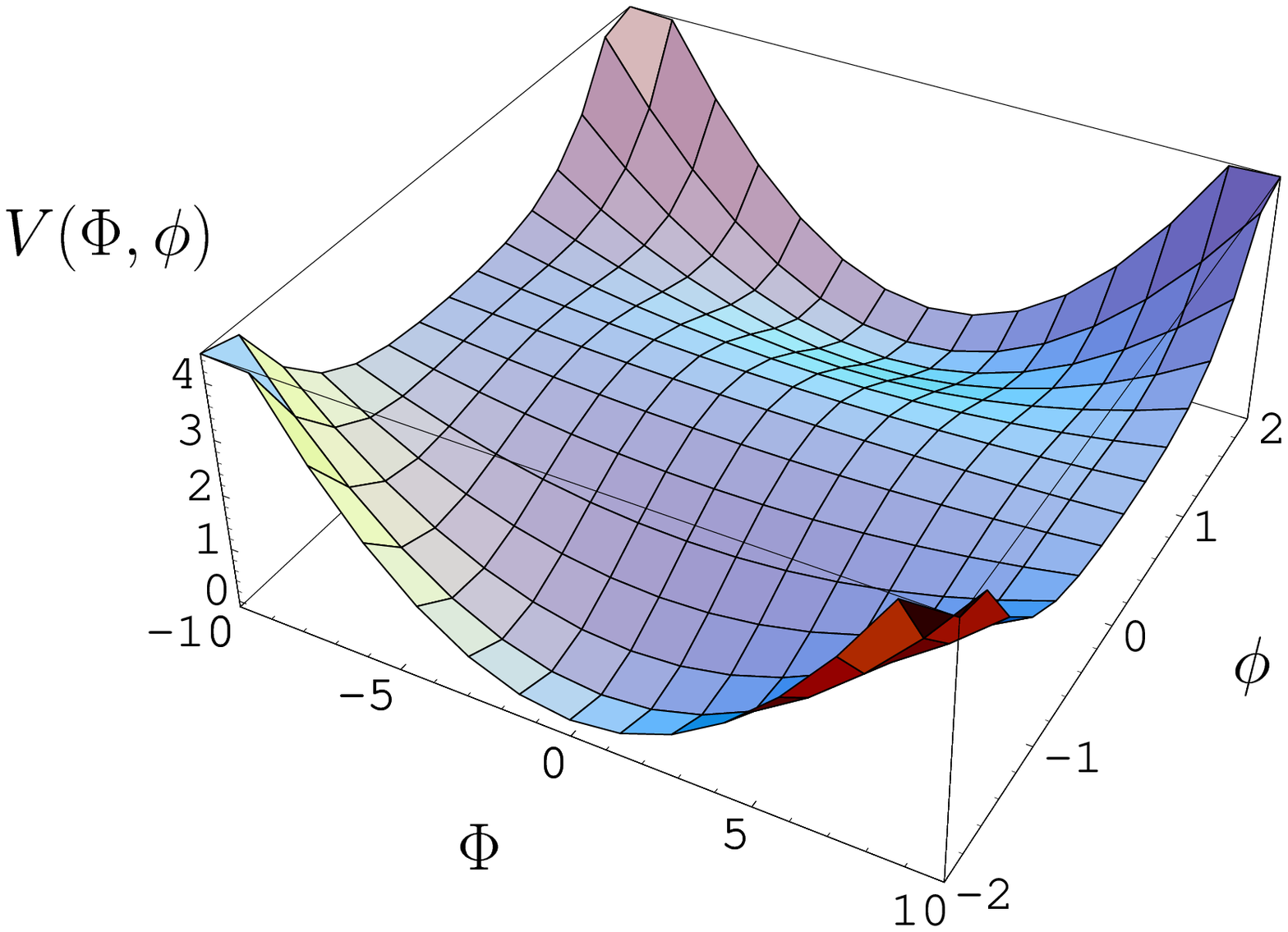}}

\vspace{-5.5cm}

\caption{
Artist's view (not in scale) of the saddle point of the potential in 
Eq.~(\ref{V}).
}
\end{center}
\end{figure}

Now, since the effective mass--squared of $\phi$ is
\begin{equation}
(m_\phi^{\rm eff})^2=\lambda\Phi^2-\alpha M^2,
\label{mphi}
\end{equation}
if \mbox{$\Phi>\Phi_c$} then $\phi$ is driven to zero,
where 
\begin{equation}
\Phi_c\equiv\sqrt{\frac{\alpha}{\lambda}}\;M\sim 
\frac{m_{3/2}}{\sqrt{\lambda}}\,.
\label{phic}
\end{equation}
Suppose, therefore, that originally the system lies in the regime, where,
\mbox{$m_{3/2}<\Phi\leq m_P$} and \mbox{$\phi\simeq 0$}. With such initial 
conditions the effective potential for $\Phi$ becomes quadratic:
\begin{equation}
V(\Phi,\phi=0)=\frac{1}{2}m_\Phi^2\Phi^2+
M_S^4.
\label{over}
\end{equation}
Since, \mbox{$\Phi<m_P$} we see that, when $\phi$ remains at the origin,  
the scalar potential is dominated by a false vacuum density corresponding to 
energy:
\begin{equation}
V_{\rm inf}^{1/4}\simeq M_S\;,
\label{Vinf}
\end{equation}
which results in a period of inflation. During this period, according to the 
Friedmann equation, we have 
\mbox{$V_{\rm inf}^{1/4}\sim\sqrt{m_PH_{\rm inf}}$}. In view of 
Eq.~(\ref{Vinf}) this means that
\begin{equation}
H_{\rm inf}\sim m_{3/2}\;.
\label{Hinf}
\end{equation}
This is why there is no slow roll, because all the masses are of the 
order of the Hubble parameter during inflation, as expected by the action of
supergravity corrections \cite{randall}. And yet, there is inflation
as long as $\phi$ remains at (or very near) the origin. 

Now, during this period the Klein-Gordon equation for $\Phi$ is
\begin{equation}
\ddot{\Phi}+3H_{\rm inf}\dot{\Phi}+m_\Phi^2\Phi=0\,,
\label{KG}
\end{equation}
where the dot denotes derivative with respect to the cosmic time $t$.
The above has a solution of the form \mbox{$\Phi\propto e^{\omega t}$},
where
\begin{equation}
\omega=-\frac{3}{2}H_{\rm inf}\left[
1\pm\sqrt{1-\frac{4}{9}\left(\frac{m_\Phi}{H_{\rm inf}}\right)^2}\;\right],
\label{wF}
\end{equation}
From Eq.~(\ref{wF}) we see that
the evolution of $\Phi$ depends on whether $m_\Phi$ is larger or not from
$\frac{3}{2}H_{\rm inf}$. We look into both cases below.

\subsection{\boldmath Fast--Roll Inflation 
($m_\Phi\leq\frac{3}{2}H_{\rm inf}$)}

Fast-roll inflation was first introduced in Ref.~\cite{fastroll}. It 
corresponds to a limited period of inflation possible when the mass of
the inflaton is comparable to the Hubble parameter, as in our case.
It turns out that, in our model, when 
\mbox{$m_\Phi\lsim\frac{3}{2}H_{\rm inf}$}, we end up with a period of 
fast-roll inflation, the details of which we will study in this section. 

In this case, as suggested by Eq.~(\ref{wF}), there are two 
exponential solutions to Eq.~(\ref{KG}), both exponentially decreasing with 
time. The solution with the positive sign corresponds to the mode which 
decreases faster and rapidly disappears. Thus, the dominant solution is
the one with the negative sign, which gives
\begin{equation}
\Phi=\Phi_0\exp(-F_\Phi\Delta N)\,,
\label{PhiN}
\end{equation}
where, \mbox{$\Delta N=H_{\rm inf}\Delta t$} is the number of the 
elapsing e-foldings and
\begin{equation}
F_\Phi\equiv\frac{3}{2}\left(1-\sqrt{1-\frac{4}{3}|\eta_\Phi|}\right)
<\frac{3}{2}\,,
\label{F}
\end{equation}
with $\eta_\Phi$ being the slow--roll parameter defined as
\begin{eqnarray}
\eta_\Phi\equiv 
\frac{m_P^2}{V}\frac{\partial^2V}{\partial\Phi^2}
 & \Rightarrow & 
|\eta_\Phi|\simeq\frac{m_\Phi^2}{3H_{\rm inf}^2}\lsim \frac{3}{4}\,,
\label{eta}
\end{eqnarray}
where 
we used that \mbox{$H_{\rm inf}^2\simeq V_{\rm inf}/3m_P^2$}.
The approximation that \mbox{$H\simeq H_{\rm inf}=$ const.} is justified
for \mbox{$\Phi_c<\Phi<m_P$} because
\begin{equation}
\epsilon\equiv-\frac{\dot{H}}{H^2}=\frac{\dot{\Phi}^2}{2m_P^2H^2}=
\frac{1}{2}F_\Phi^2\left(\frac{\Phi}{m_P}\right)^2\ll 1\,,
\label{eps}
\end{equation}
where we have used Eq.~(\ref{KG}) and also that 
\mbox{$3(m_PH)^2=\frac{1}{2}\dot{\Phi}^2+V(\Phi)$} with 
\mbox{$\dot{\Phi}=-F_\Phi H_{\rm inf}\Phi$}, according to Eq.~(\ref{PhiN}).

Therefore, in view of Eq.~(\ref{PhiN}), we find that the total number of
e-foldings corresponding to fast-roll inflation is given by
\begin{equation}
N_{\rm FR}=-\frac{1}{F_\Phi}\ln\left(\frac{\Phi_{\rm end}}{\Phi_0}\right)
\simeq \frac{1}{F_\Phi}\ln\left(\frac{m_P}{m_{3/2}}\right)
+\frac{1}{F_\Phi}\ln\sqrt{\lambda}\,,
\label{Nfr1}
\end{equation}
where \mbox{$\Phi_0\sim m_P$} and 
\mbox{$\Phi_{\rm end}=\Phi_c\sim m_{3/2}/\sqrt{\lambda}$} 
are the initial and final values for the roll of $\Phi$ respectively
[cf. Eq.~(\ref{phic})]. From the above it is evident that the larger 
$m_\Phi$ is the larger $F_\Phi$ is and, therefore, the smaller the number 
$N_{\rm FR}$ of the total e-foldings of Fast-Roll inflation.\footnote{In the 
extreme limit, when \mbox{$m_\Phi\ll H_{\rm inf}$} we obtain 
\mbox{$F_\Phi\rightarrow|\eta_\Phi|$} and we have the usual slow--roll
inflation.} However, this number 
cannot become arbitrarily small because, if $m_\Phi$ is bigger than
$\frac{3}{2}H_{\rm inf}$ then the dynamics of $\Phi$ becomes distinctly 
different. We explore this case in what follows.

\subsection{\boldmath Locked Inflation ($m_\Phi>\frac{3}{2}H_{\rm inf}$)}

Locked inflation was introduced in Ref.~\cite{dvali}, using a potential of
the form shown in Eq.~(\ref{V}). This kind of inflation uses
an inflaton field, which is oscillating on top of the false vacuum density
responsible for inflation.\footnote{Inflation with an oscillating inflaton,
but without an additional false vacuum contribution, has also been studied
in \cite{damour} in the context of non--convex potentials. Unfortunately it 
has been found that such inflation can last no more than about 10 e-foldings.} 
It turns out that, in our model, in the case when 
\mbox{$m_\Phi>\frac{3}{2}H_{\rm inf}$} we obtain this kind of behaviour for
$\Phi$. 

Indeed, in this case the Klein-Gordon Eq.~(\ref{KG}) is solved by an equation 
of the form 
\begin{equation}
\Phi=\bar{\Phi}(\Delta t)\,\cos(\omega_\Phi\Delta t)\,,
\label{Phiosc}
\end{equation}
where 
\begin{equation}
\omega_\Phi=H_{\rm inf}
\sqrt{\left(\frac{m_\Phi}{H_{\rm inf}}\right)^2-\frac{9}{4}}
\;\approx m_\Phi\sim m_{3/2}
\label{w}
\end{equation}
and
\begin{equation}
\bar{\Phi}=\bar{\Phi}_0\exp\left(-\frac{3}{2}\,\Delta N\right)\,.
\label{barphieq}
\end{equation}

It may strike as odd that the field is oscillating instead of rolling toward
the true minimum of the system. This is because, provided the frequency of the 
oscillations is large enough, the time that the oscillating field spends on 
top of the saddle point of the potential is too small to allow its escape from 
the oscillatory trajectory. Indeed, as shown in Eq.~(\ref{w}), the 
oscillation frequency is \mbox{$\omega_\Phi\sim m_\Phi$} and the time 
interval that the field spends on top of the saddle point 
(\mbox{$\Delta\Phi\leq\Phi_c$}) is
\begin{eqnarray}
\omega_\Phi\Delta t\sim\frac{\Delta\Phi}{\bar{\Phi}} & \Rightarrow & 
(\Delta t)_{\rm saddle}\sim\frac{\Phi_c}{m_\Phi\bar{\Phi}}\sim
\frac{1}{\sqrt{\lambda}}
\frac{1}{\bar{\Phi}}\,,
\label{Dt}
\end{eqnarray}
where $\bar{\Phi}$ is the amplitude of the oscillations. Originally this 
amplitude may be quite large \mbox{$\bar{\Phi}\sim m_P$} but the expansion of 
the Universe dilutes the energy of the oscillations and, therefore, 
$\bar{\Phi}$ decreases, which means that $(\Delta t)_{\rm saddle}$ grows. 
However, until $(\Delta t)_{\rm saddle}$ becomes large enough to be comparable 
to the inverse of the tachyonic mass of $\phi$, the latter has no time to roll
away from the saddle. Hence, the oscillations of $\Phi$ on top of the saddle
continue until the amplitude decreases down to
\begin{equation}
\bar{\Phi}_{\rm end}\sim m_{3/2}/\sqrt{\lambda}\sim\Phi_c\;,
\label{Phiend}
\end{equation}
at which point $\phi$ departs from the origin and rolls down toward its 
vacuum expectation value (VEV) $M$. 
During the oscillations the density of the oscillating $\Phi$ is 
\begin{equation}
\rho_\Phi=\frac{1}{2}\dot{\Phi}^2+\frac{1}{2}m_\Phi^2\Phi^2\simeq
\frac{1}{2}m_\Phi^2\bar{\Phi}^2.
\label{rphi}
\end{equation}
Comparing this with the overall potential density given in Eq.~(\ref{over}) we 
see that, for oscillation amplitude smaller than $m_P$, the overall density is 
dominated by the false vacuum density given in Eq.~(\ref{Vinf}), which remains 
constant as long as $\phi$ remains locked at the origin. Hence, the Universe 
undergoes a period of inflation when $\bar{\Phi}$ lies in the region
\begin{equation}
\bar{\Phi}\in \left(m_{3/2}/\sqrt{\lambda},\;m_P\right)\,. 
\label{range}
\end{equation}

Therefore, in view of Eq.~(\ref{barphieq}), the total number of e-foldings of 
locked inflation corresponds to the range shown in Eq.~(\ref{range}) and is 
given by
\begin{equation}
N_{\rm lock}=\frac{2}{3}\ln\left(\frac{m_P}{m_{3/2}}\right)
+\frac{2}{3}\ln\sqrt{\lambda}
\leq 24
\,.
\label{Nlock}
\end{equation}
\begin{center}
\begin{figure}
\leavevmode
\hbox{%
\epsfxsize=5in
\epsffile{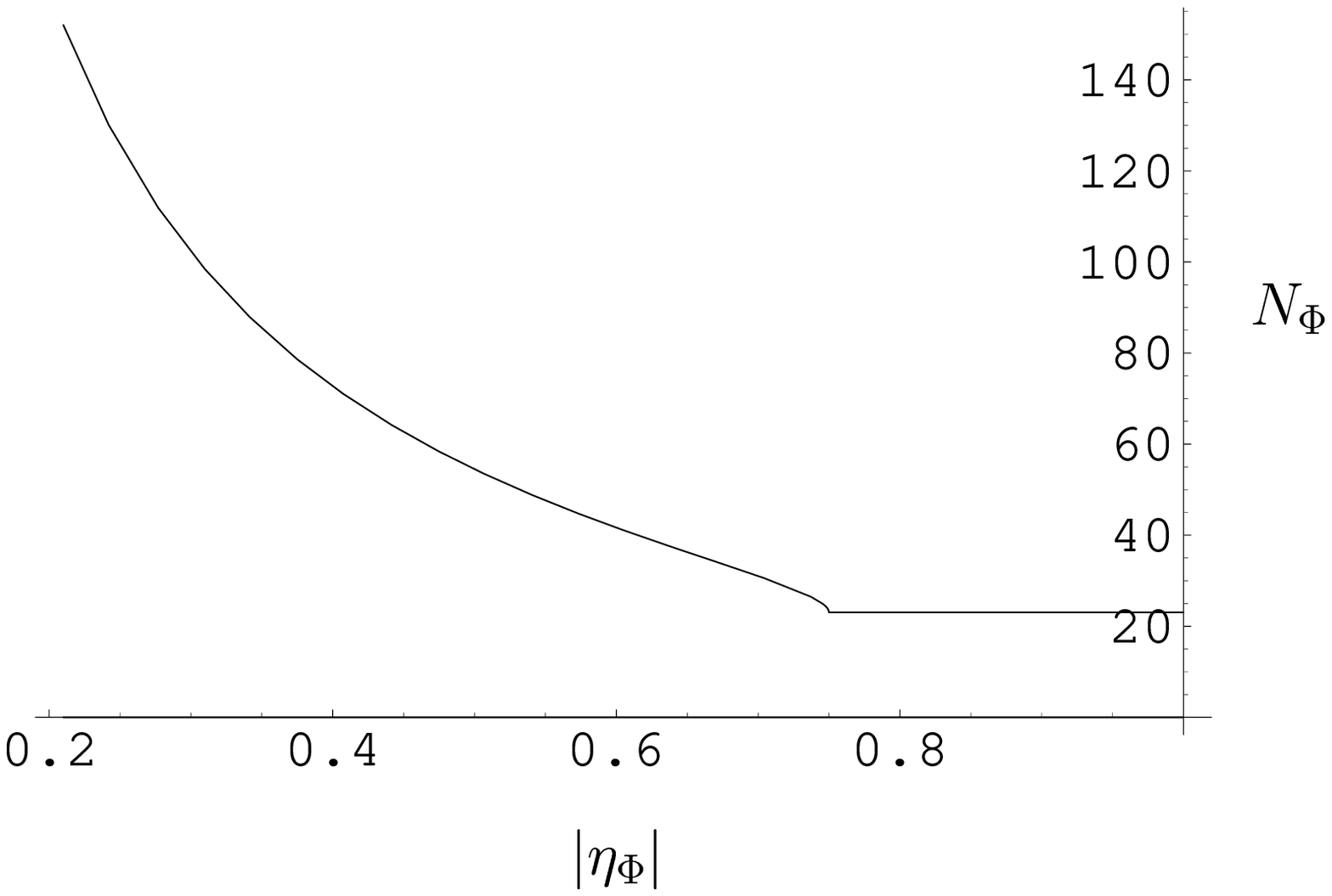}}
\vspace{-9cm}
\caption{
Plot of the number of e-foldings $N_\Phi$ (corresponding to 
inflation while \mbox{$\phi\approx 0$}) against the curvature of the potential
along the direction of $\Phi$ (parameterised by the slow-roll parameter
$\eta_\Phi$) near the transition point between fast-roll 
(\mbox{$|\eta_\Phi|<3/4$}) and locked (\mbox{$|\eta_\Phi|\geq 3/2$}) inflation.
As shown $N_\Phi$ decreases with increasing $|\eta_\Phi|$ until the 
critical value \mbox{$|\eta_\Phi|=3/2$} is reached, above which it remains 
constant; \mbox{$N_\Phi=N_{\rm lock}$}. For illustrative purposes, 
we have chosen \mbox{$\lambda=1$}.
}
\end{figure}
\end{center}
From the above and in view also of Eqs.~(\ref{F}) and (\ref{Nfr1}) we see that
\mbox{$N_{\rm FR}>N_{\rm lock}$}. Hence, we have shown that {\em $N_{\rm lock}$
is the minimum number of e-foldings that the Universe inflates while $\phi$ 
remains at (or very near) the origin}. Thus, even though the slope for the 
rolling $\Phi$ field maybe arbitrarily steep, locked inflation guarantees that 
there are at least a fixed number of e-foldings of inflationary expansion. In 
Figure~2 we plot $N_\Phi$ in the neighbourhood of the transition
between fast-roll and locked inflation.

However, from Eq.~(\ref{Nlock}), we see that, for realistic values of the 
parameters, locked inflation alone cannot provide the necessary number of 
e-foldings corresponding to the cosmological scales. Fortunately, there is a 
subsequent period of inflation, this time driven by the scalar field $\phi$, 
after it departs from zero and rolls toward its VEV. This is another period of 
fast-roll inflation and we discuss it next.

\section{Tachyonic Fast--Roll Inflation}

In Ref.~\cite{fastroll}, fast-roll inflation corresponds to the roll of a 
field $\phi$ from a local maximum of its potential, when its tachyonic mass is 
comparable to the Hubble parameter, which is exactly the case for the $\phi$ 
field in our model. The $\phi$ field corresponds to the so--called waterfall 
field in regular hybrid inflation, which is thought to cause a phase 
transition that terminates inflation.\footnote{The second stage of fast-roll 
inflation in non-flat hybrid inflation can be also viewed as a slowly 
progressing phase transition, which terminates hybrid inflation. In such 
manner it has been studied in Ref.~\cite{anupam}.} However, in our case, 
inflation continues after the phase transition. 

The tachyonic potential for $\phi$ is of the form [cf. Eq.~(\ref{V})]
\begin{equation}
V(\phi)=V_{\rm inf}-\frac{1}{2}|(m_\phi^{\rm eff})^2|\phi^2+
\frac{1}{4}\alpha\phi^4,
\label{Vtachy}
\end{equation}
where \mbox{$V_{\rm inf}=M_S^4$} and
$(m_\phi^{\rm eff})^2$ is given by Eq.~(\ref{mphi}). Since the roll of $\phi$
begins after \mbox{$\bar{\Phi}<\Phi_c\sim m_{3/2}/\sqrt{\lambda}$}, we see that
\mbox{$|(m_\phi^{\rm eff})^2|\sim m_{3/2}^2$} and also 
\mbox{$(m_\phi^{\rm eff})^2\simeq -m_\phi^2$}. 

Now, the Klein-Gordon satisfied by $\phi$ is
\begin{equation}
\ddot{\phi}+3H_{\rm inf}\dot{\phi}-m_\phi^2\phi=0\,,
\label{KG1}
\end{equation}
where \mbox{$m_\phi\sim m_{3/2}$} as shown in Eq.~(\ref{mphi0}). The above 
admits solutions of the form \mbox{$\phi\propto\exp(\omega_\phi\Delta t)$} with
\begin{equation}
\omega_\phi=-\frac{3}{2}\;H_{\rm inf}
\left[1\pm\sqrt{1+\frac{4}{9}\left(\frac{m_\phi}{H_{\rm inf}}\right)^2}\;\right]
\label{wphi}
\end{equation}
The solution with the positive sign corresponds to the exponentially decreasing
mode which rapidly disappears, whereas the solution with the negative sign
corresponds to the exponentially growing mode:
\begin{equation}
\phi=\phi_0\exp(F_\phi\Delta N)\,,
\label{phiN}
\end{equation}
where, \mbox{$\Delta N=H_{\rm inf}\Delta t$} is the number of the 
elapsing e-foldings and
\begin{equation}
F_\phi\equiv\frac{3}{2}\left(\sqrt{1+\frac{4}{3}|\eta_\phi|}-1\right)\,,
\label{F1}
\end{equation}
with $\eta_\phi$ being the slow--roll parameter defined in a similar manner as
in Eq.~(\ref{eta}). For $\eta_\phi$ we have
\begin{eqnarray}
|\eta_\phi|\equiv 
\frac{m_P^2}{V}\left|\frac{\partial^2V}{\partial\phi^2}\right|
\simeq\frac{m_\phi^2}{3H_{\rm inf}^2}\sim 1\,,
\label{eta1}
\end{eqnarray}
The approximation that \mbox{$H\simeq H_{\rm inf}=$ const.} is again justified
for \mbox{$0<\phi<M\sim m_P$} because
\begin{equation}
\epsilon=\frac{1}{2}F_\phi^2\left(\frac{\phi}{m_P}\right)^2\ll 1\,,
\label{eps1}
\end{equation}
where we have used Eq.~(\ref{KG1}) and also that 
\mbox{$3(m_PH)^2=\frac{1}{2}\dot{\phi}^2+V(\phi)$} with 
\mbox{$\dot{\phi}=F_\phi H_{\rm inf}\phi$}
[cf. Eq.~(\ref{phiN})].

Therefore, in view of Eq.~(\ref{phiN}), we find that the total number of
e-foldings corresponding to fast-roll inflation is given by
\begin{equation}
N_\phi=\frac{1}{F_\phi}\ln\left(\frac{\phi_{\rm end}}{\phi_0}\right)
\simeq \frac{1}{F_\phi}\ln\left(\frac{M}{m_\phi}\right)\,,
\label{Nfr2}
\end{equation}
where $\phi_0$ and $\phi_{\rm end}$ are the initial and final values for the 
roll of $\phi$. The final value of $\phi$ is its VEV \mbox{$M\sim m_P$}, while 
the initial value of $\phi$ depends on the initial conditions at the onset of 
inflation (see Sec.~\ref{init} for a more detailed discussion on this issue). 
If $\phi$ is very close to the origin then $\phi_0$ is determined by the 
tachyonic fluctuations which send it off the top of the potential, and is given
by \mbox{$\phi_0=m_\phi/2\pi$} \cite{felder}. This is what we assume for the 
moment in Eq.~(\ref{Nfr2}). 

From Eqs.~(\ref{Nfr1}), (\ref{Nlock}) and (\ref{Nfr2}) we see that the total 
number of inflationary e-foldings is given by 
\begin{equation}
N_{\rm tot}=N_\Phi+N_\phi\simeq 
\left(\frac{1}{\bar F_\Phi}+\frac{1}{F_\phi}\right)
\ln\left(\frac{m_P}{m_{3/2}}\right)
+\frac{1}{\bar F_\Phi}\ln\sqrt{\lambda}
\,,
\label{Ntot}
\end{equation}
where we have set 
\begin{equation}
\bar F_\Phi\equiv\min\left\{\frac{3}{2},\,F_\Phi\right\}\,
\label{barF}
\end{equation}
and $N_\Phi$ corresponds to the first stage of inflation and is given by 
[cf. Eqs.~(\ref{Nfr1}) and (\ref{Nlock})]
\begin{equation}
N_\Phi=
\simeq \frac{1}{\bar F_\Phi}\ln\left(\frac{m_P}{m_{3/2}}\right)
+\frac{1}{\bar F_\Phi}\ln\sqrt{\lambda}\,,
\label{NPhi}
\end{equation}

It is the above number $N_{\rm tot}$, which needs to be compared to the 
necessary e-foldings for the cosmological scales.

\section{The necessary e-foldings}

Inflation solves in a single stroke the horizon and flatness problems of the
Standard Hot Big Bang (SHBB) cosmology, while providing also the superhorizon 
spectrum of density perturbations necessary for the formation of Large Scale 
Structure. To do all these, the inflationary period has to be sufficiently
long because the scales that correspond to the cosmological observations need 
to exit the horizon during inflation. The largest of these scales is determined
by the requirements of the horizon problem and corresponds to about 100 times 
the scale of the present Horizon. The number of e-foldings required to inflate 
this scale on superhorizon size provides a lower bound on the total number of
e-foldings of inflation and it is estimated as follows \cite{book}:
\begin{equation}
N_{\rm cosmo}=72-\ln\left(\frac{m_P}{V_{\rm inf}^{1/4}}\right)-
\frac{1}{3}\ln\left(\frac{V_{\rm inf}^{1/4}}{T_{\rm reh}}\right)-N_0\;,
\label{N}
\end{equation}
where $V_{\rm inf}^{1/4}$ is the energy scale of inflation, $T_{\rm reh}$ is 
the reheat temperature, corresponding to the temperature of the thermal bath 
when the SHBB begins after the entropy production at the end of inflation, and 
\mbox{$N_0\geq 0$} is the total e-foldings that correspond to any subsequent 
periods of inflation. The reheat temperature is given by
\begin{equation}
T_{\rm reh}\sim\sqrt{\Gamma m_P}\;,
\label{Treh}
\end{equation}
where
\begin{equation}
\Gamma\simeq g^2m_\phi
\label{G}
\end{equation}
is the decay rate for the inflaton field corresponding to the last stage of
inflation and $g$ is the coupling of $\phi$ to the decay products. If the
coupling of the field to other particles is extremely weak then the field
will decay predominantly through gravitational couplings, in which case
\mbox{$\Gamma\sim m_\phi^3/m_P^2$}. Thus, the effective range for $g$ is
\begin{equation}
\frac{m_{3/2}}{m_P}\leq g\leq 1\,,
\label{grange}
\end{equation}
where we considered that \mbox{$m_\phi\sim m_{3/2}$}.

Combining Eqs.~(\ref{Vinf}), (\ref{Treh}) and (\ref{G}) we can recast 
Eq.~(\ref{N}) as
\begin{equation}
N_{\rm cosmo}=
72-\frac{1}{2}\ln\left(\frac{m_P}{m_{3/2}}\right)+\frac{1}{3}\ln g-N_0\simeq
54-\frac{1}{3}\ln g-N_0
\label{Ncosmo}
\end{equation}

If our model is to explain the cosmological observations we have to demand 
that \mbox{$N_{\rm tot}>N_{\rm cosmo}$}. This provides an upper bound on 
$m_\phi$. Indeed, after a little algebra, it can be shown that 
Eqs.~(\ref{Ntot}), (\ref{NPhi}) and (\ref{Ncosmo}), in view also 
of Eqs.~(\ref{F1}) and (\ref{eta1}), imply the bound
\begin{equation}
\frac{m_\phi}{H_{\rm inf}}<\frac{3}{2}
\left\{\left[\left(
\frac{108+\ln\sqrt{g}-\frac{3}{2\bar F_\Phi}\ln\sqrt{\lambda}
-\frac{3}{2}N_0}{\ln(m_P/m_{3/2})}
-\frac{3}{4}\left(1+\frac{2}{\bar F_\Phi}\right)
\right)^{-1}+1\right]^2-1\right\}^{1/2},
\label{bound}
\end{equation}
which is more stringent the smaller $N_\Phi$ is. Thus, the 
tightest bound corresponds to a first stage of locked inflation, where
\mbox{$N_\Phi=N_{\rm lock}$} or, equivalently, when \mbox{$\bar F_\Phi=3/2$}. 
In this case the above bound becomes:
\begin{equation}
\frac{m_\phi}{H_{\rm inf}}<\frac{3}{2}
\left\{\left[\left(
\frac{108+\ln\sqrt{g/\lambda}-\frac{3}{2}N_0}{\ln(m_P/m_{3/2})}-\frac{7}{4}
\right)^{-1}+1\right]^2-1\right\}^{1/2}
\label{lockbound}
\end{equation}

Clearly, the above suggest that the upper bound on $m_\phi/H_{\rm inf}$ is of 
order unity. For example, for the range given in Eq.~(\ref{grange}) and if we 
choose \mbox{$N_0=0$} and \mbox{$\lambda\sim 1$} it is easy to show that the 
above bound interpolates between 2 and 3. Moreover, if the mass of $\Phi$ is 
below $\frac{3}{2}H_{\rm inf}$ then \mbox{$N_\Phi=N_{\rm FR}\geq N_{\rm lock}$}
and the bound on $m_\phi$ is further relaxed because $N_\phi$ does not need to 
be as large as before. 

Consequently, it seems that, regardless of $m_\Phi$, the required e-foldings 
of inflation corresponding to the cosmological scales can be attained, only 
with a mild upper bound on $m_\phi$. However, there is one grave danger that 
we had overlooked and this is the possibility of disastrous Primordial Black 
Hole production due to the phase transition that terminates the first stage of 
inflation. We elaborate on this problem in the next section.

\section{The danger from Primordial Black Hole production}

A rather important issue to be investigated is the possibility of excessive
Primordial Black Hole (PBH) production after the end of inflation, when the
scale, which corresponds to the phase transition that releases $\phi$ from
the top of the saddle, reenters the horizon.

\subsection{The PBH calamity}

It is well known that the outburst of tachyonic fluctuations at the phase 
transition can generate a mountain of density/curvature perturbations 
(localised around the scale corresponding to the phase transition)
with amplitude of order unity \cite{kofman}. When 
these perturbations reenter the horizon and become causally connected they can 
collapse and form PBHs \cite{carr}. 
The mass of these PBHs is of the order of the mass 
included in the horizon volume at the time of reentry, i.e.
\begin{equation}
M_{\sc pbh}\sim\left.\frac{\rho}{H^3}\right|_{\sc pbh}
\sim\frac{m_P^2}{H_{\sc pbh}}\,,
\label{Mpbh}
\end{equation}
where the subscript `{\sc pbh}' denotes the epoch of PBH formation. The 
probability of PBH formation is of order unity, which means that a sizable 
fraction of the energy density of the Universe collapses into PBHs and then 
scales as pressureless matter. Hence, just after their formation, the PBHs 
dominate the density of the Universe and result in a period of matter 
domination. This period lasts until the PBHs evaporate, which occurs
after time $\Delta t_{\rm ev}$, where \cite{hawk}
\begin{equation}
\Delta t_{\rm ev}\sim\frac{M_{\sc pbh}^3}{m_P^4}\;.
\label{Dtev}
\end{equation}

One of the greatest successes of the SHBB is the correct prediction
for the delicate abundance of the light elements. They are generated during a 
process called Big Bag Nucleosynthesis (BBN), which takes place
at temperatures \mbox{$T_{\sc bbn}\sim 1$ MeV}, at cosmic time of about 1~sec.
The BBN process is very sensitive to the state of the Universe at the time.
Consequently, it is imperative that the SHBB has begun before BBN occurs.

Therefore, the PBHs must evaporate before BBN, i.e. 
\mbox{$\Delta t_{\rm ev}<H_{\sc bbn}^{-1}\sim 1$ sec}. This requirement results
in a constraint on $M_{\sc pbh}$, which reads
\begin{equation}
M_{\sc pbh}^3<\frac{m_P^5}{T_{\sc bbn}^2}\qquad\Rightarrow\qquad
M_{\sc pbh}<10^{32}{\rm GeV}\sim 10^8{\rm g}\,.
\label{Mpbhbound}
\end{equation}
To calculate the value of $M_{\sc pbh}$ we use the fact that the formation of
the PBHs occurs when the scale that corresponds to the phase transition, 
reenters the horizon. Since this scale exits the horizon $N_\phi$ e-foldings
before the end of inflation, it is easy to find
\begin{equation}
\frac{a_{\sc pbh}}{a_{\rm inf}}\sim\exp(N_\phi)
\sim\left(\frac{m_P}{m_\phi}\right)^{1/F_\phi},
\label{afrac}
\end{equation}
where $a(t)$ is the scale factor of the Universe, the subscript `inf' here
denotes the end of inflation and we used Eq.~(\ref{Nfr2}). To proceed we need 
to consider individually the cases when the PBH formation takes place before 
and after reheating.

\subsubsection{\boldmath PBH formation before reheating 
($H_{\sc pbh}\geq\Gamma$)}

In this case the Universe after the end of inflation and until the formation of
PBHs remains matter dominated so that \mbox{$a\propto H^{-2/3}$}. 
Then, Eq.~(\ref{afrac}) gives
\begin{equation}
H_{\sc pbh}\sim H_{\rm inf}
\left(\frac{m_P}{m_\phi}\right)^{-3/2F_\phi}.
\label{Hpbh1}
\end{equation}
Substituting the above into Eq.~(\ref{Mpbh}) and considering also that
\mbox{$m_\phi\sim m_{3/2}\sim H_{\rm inf}$} we obtain
\begin{equation}
M_{\sc pbh}\sim m_P\left(\frac{m_P}{m_{3/2}}\right)^{\frac{3}{2F_\phi}+1}.
\label{Mpbh1}
\end{equation}
Enforcing in the above the bound in Eq.~(\ref{Mpbhbound}) we end up with the
requirement
\begin{equation}
\left(\frac{m_P}{m_{3/2}}\right)^{\frac{9}{2F_\phi}+1}<
\left(\frac{m_{3/2}}{T_{\sc bbn}}\right)^2,
\label{lethal}
\end{equation}
which is impossible to satisfy with \mbox{$F_\phi\geq 0$}. 

Note that, if \mbox{$F_\phi\gg 1$} then Eq.~(\ref{lethal}) can be recast as
\mbox{$V_{\rm inf}^{1/4}>m_P(T_{\sc bbn}/m_P)^{1/3}$}, where we used
Eq.~(\ref{Vinf}). Satisfying this bound is 
marginally viable but requires a huge value of $F_\phi$ or, equivalently, 
\mbox{$m_\phi\gg H_{\rm inf}$}, which eliminates the second stage of 
inflation.

\subsubsection{\boldmath PBH formation after reheating ($H_{\sc pbh}<\Gamma$)}

In this case, after the end of inflation, the Universe is matter dominated 
(\mbox{$a\propto H^{-2/3}$}) until reheating, but afterwards and until PBH 
formation it becomes radiation dominated with \mbox{$a\propto H^{-1/2}$}. 
Therefore, Eq.~(\ref{afrac}) gives
\begin{equation}
H_{\sc pbh}\sim H_{\rm inf}
\left(\frac{H_{\rm inf}}{\Gamma}\right)^{1/3}
\left(\frac{m_P}{m_\phi}\right)^{-2/F_\phi}.
\label{Hpbh2}
\end{equation}
Substituting the above into Eq.~(\ref{Mpbh}) and considering also that
\mbox{$m_\phi\sim m_{3/2}\sim H_{\rm inf}$} we obtain
\begin{equation}
M_{\sc pbh}\sim g^{2/3}m_P
\left(\frac{m_P}{m_{3/2}}\right)^{\frac{2}{F_\phi}+1},
\label{Mpbh2}
\end{equation}
where we also used Eq.~(\ref{G}).
Enforcing in the above the bound in Eq.~(\ref{Mpbhbound}) we end up with the
requirement
\begin{equation}
g^2\left(\frac{m_P}{m_{3/2}}\right)^{\frac{6}{F_\phi}+1}<
\left(\frac{m_{3/2}}{T_{\sc bbn}}\right)^2.
\label{gbound}
\end{equation}
Now, it is evident that Eq.~(\ref{Hpbh2}) can be recast as
\begin{equation}
\frac{H_{\sc pbh}}{\Gamma}\sim g^{-8/3}
\left(\frac{m_P}{m_{3/2}}\right)^{-2/F_\phi}.
\label{gcond}
\end{equation}
Enforcing into the above the bound in Eq.~(\ref{gbound}) and taking also into
account that \mbox{$H_{\sc pbh}<\Gamma$} it is easy to show that we end up 
again with Eq.~(\ref{lethal}). 

Therefore, it seems that, in our model, PBH production unavoidably disturbs 
BBN, since it turns out that the PBHs cannot evaporate early enough. This 
disappointing result has been reached numerically also in Ref.~\cite{bastards}.
The only escape from this catastrophe is to avoid producing any PBHs in the 
first place. In contrast to Ref.~\cite{bastards}, we show that there is ample
parameter space where this can be achieved in a natural way. The key to the 
solution is to avoid fixing \mbox{$\phi\simeq 0$} as the initial condition 
for $\phi$. Indeed, in the following section we elaborate on this issue and 
consider a more realistic initial value for $\phi$, which, for small enough 
$\lambda$, results in no PBH production.

\subsection{The solution to the PBH problem}

\subsubsection{A way out}

One can avoid the generation of PBHs if, at the time when the amplitude of the 
oscillating $\Phi$ is decreased down to $\Phi_c$, the field $\phi$ is 
significantly displaced from the top of the saddle so that the tachyonic 
fluctuations are suppressed. Hence, we require that \mbox{$\phi_0>m_\phi$}.
Writing 
\begin{equation}
\phi_0\equiv\beta m_\phi\;,
\label{b}
\end{equation}
to avoid PBH production we need
\begin{equation}
1\ll\beta\ll\frac{M}{m_\phi}\sim 10^{15}\,,
\label{brange}
\end{equation}
where the upper bound ensures the `locking' of $\phi$ on top of the saddle.
In view of the above, Eq.~(\ref{Nfr2}) becomes
\begin{equation}
N_\phi\simeq\frac{1}{F_\phi}\ln\left(\frac{M}{m_\phi}\right)-
\frac{1}{F_\phi}\ln\beta
\label{newN}
\end{equation}
Using this we can reproduce the bound in Eq.~(\ref{bound}), which now becomes
\begin{eqnarray}
\frac{m_\phi}{H_{\rm inf}} & < & \frac{3}{2}
\left\{\left[\left(
\frac{108+\ln\sqrt{g}-\frac{3}{2\bar F_\Phi}\ln\sqrt{\lambda}
-\frac{3}{2}N_0}{\ln(m_P/m_{3/2})}
-\frac{3}{4}\left(1+\frac{2}{\bar F_\Phi}\right)
\right)^{-1}\right.\right.\times\nonumber\\
& & \times\left.\left.
\left(1-\frac{\ln\beta}{\ln(m_P/m_{3/2})}\right)
+1\right]^2-1\right\}^{1/2},
\label{newbound}
\end{eqnarray}
%
where, from Eq.~(\ref{brange}), it is evident that
\begin{equation}
0<\frac{\ln\beta}{\ln(m_P/m_{3/2})}<1\,.
\label{lnbrange}
\end{equation}
Thus, we see that the upper bound on the tachyonic mass of the $\phi$ modulus 
is somewhat strengthened. 

\subsubsection{\boldmath The initial value of $\phi$ at the phase transition}
\label{init}

To obtain an estimate of the likely value of $\beta$ we have to take a closer
look on the evolution of $\phi$ during the first period of inflation, when 
\mbox{$\bar{\Phi}\gg\Phi_c$}. Then we can write the Klein-Gordon equation of 
$\phi$ as
\begin{equation}
\dot{\phi}\ddot{\phi}+3H_{\rm inf}\dot{\phi}^2+
V'(\phi)\dot{\phi}= 0\,,
\label{newKG}
\end{equation}
where the prime here denotes derivative with respect to $\phi$ and we have
multiplied the Klein-Gordon with $\dot\phi$. Now, when \mbox{$\bar\Phi>\Phi_c$}
we have \mbox{$(m_\phi^{\rm eff})^2\simeq\lambda\Phi^2$} and\footnote{The 
quartic term $\sim\alpha\phi^4$ in the scalar potential in Eq.~(\ref{V}) is 
important only for \mbox{$\phi\gsim M\sim m_P$}.}
\mbox{$V(\phi)\simeq\frac{1}{2}\lambda\Phi^2\phi^2$}. Using this it is 
easy to show that
\begin{equation}
V'(\phi)\dot\phi=\dot V(\phi)-2V(\phi)(\dot\Phi/\Phi)\simeq
\dot V(\phi)+3H_{\rm inf}V(\phi)\,,
\label{dotV}
\end{equation}
where we have approximated \mbox{$\Phi\sim\bar\Phi$} (which holds during 
almost the entire oscillation period of $\Phi$)\footnote{We consider only the 
envelope of the oscillating $\Phi$. The only effect of the oscillations 
themselves is to result, possibly, in non-perturbative production 
(via parametric resonance) of $\phi$--particles, which may take place during 
each period when \mbox{$\lambda\Phi(t)^2<m_\phi^2$} even though 
\mbox{$\bar\Phi\gg\Phi_c$}. However, as in preheating \cite{preh}, this 
particle production removes only a fraction of the energy of the oscillations 
and, therefore, does not significantly back-react to the oscillating 
zero mode of $\Phi$. Since we attempt only order-of-magnitude estimates here, 
we can safely neglect this effect. Note, also, that the perturbative decay of
$\Phi$ into $\phi$ is possible only if 
\mbox{$m_\Phi>2|m_\phi^{\rm eff}|\sim\sqrt{\lambda}\bar\Phi$}, which cannot 
occur for \mbox{$\bar\Phi>\Phi_c$}. Finally, because of the absence of a 
quartic term $\sim\Phi^4$ in the scalar potential in Eq.~(\ref{V}), the decay 
of the $\Phi$ zero mode into $\Phi$--particles of higher momenta is also 
suppressed.}
and considered that [cf. Eq.~(\ref{rphi})] 
\mbox{$\rho_\Phi\propto\bar\Phi^2\propto\exp(-3H_{\rm inf}t)$}, which
results in \mbox{$\dot\Phi/\Phi\sim-\frac{3}{2}H_{\rm inf}$}.\footnote{
Note that $\Phi$ oscillates in a quadratic potential 
\mbox{$V(\Phi)\sim m_\Phi^2\Phi^2$} and, therefore, it corresponds to a 
collection of massive $\Phi$--particles, which behave like pressureless matter
\cite{book1}\cite{KT}. This means that \mbox{$\rho_\Phi\propto a^{-3}$}, where
the scale factor of the Universe, during inflation, is
\mbox{$a(t)\propto\exp(-H_{\rm inf}t)$}.}

Using the above, Eq.~(\ref{newKG}) can be written as
\begin{equation}
\dot{\rho}_\phi+3H_{\rm inf}[2\rho_\phi^{\rm kin}+V(\phi)]=0\,,
\label{rhoKG}
\end{equation}
where the kinetic density of $\phi$ is 
\mbox{$\rho_\phi^{\rm kin}\equiv\frac{1}{2}\dot\phi^2$} and 
\mbox{$\rho_\phi=\rho_\phi^{\rm kin}+V(\phi)$}. Now, since during the
first period of inflation we have 
\mbox{$(m_\phi^{\rm eff})^2\sim
\lambda\bar\Phi^2\gg\lambda\Phi_c^2\sim H_{\rm inf}$}, during a Hubble time
$\phi$ undergoes a huge number of oscillations. This means that these 
oscillations are, to a very good approximation, harmonic. Consequently,
following the reasoning of \cite{KT} (see also \cite{book1}), we can consider 
\mbox{$\bar\rho_\phi^{\rm kin}\simeq\bar V(\phi)\simeq\frac{1}{2}\rho_\phi$},
where the bar here means ``average per oscillation''. Hence, Eq.~(\ref{rhoKG})
can be recast as
\begin{equation}
\dot{\rho}_\phi+\frac{9}{2}H_{\rm inf}\rho_\phi\simeq 0\,.
\label{rKG}
\end{equation}
This means that 
\mbox{$\rho_\phi\simeq 2\bar V(\phi)\propto \bar\Phi^2\bar\phi^2\propto
\exp(-\frac{9}{2}H_{\rm inf}t)$}. Hence, we see that 
\mbox{$\rho_\phi/\rho_\Phi\propto\exp(-\frac{3}{2}H_{\rm inf}t)$}, which means 
that the evolution of $\phi$ cannot influence the dynamics of the first stage 
of inflation (because soon after the onset of inflation $\rho_\phi$ becomes 
negligible compared to $\rho_\Phi$).

Now, considering that \mbox{$\bar\Phi^2\propto\exp(-3H_{\rm inf}t)$} we see
that \mbox{$\bar\phi\propto\exp(-\frac{3}{4}H_{\rm inf}t)$}. Hence, for the 
typical value 
of $\phi$ at the end of the first stage of inflation we have found
\begin{equation}
\phi_0\sim\bar\phi_0\sim\phi_{\rm in}\,\exp\left(-\frac{3}{4}N_\Phi\right)
\label{f0}
\end{equation}
where $\phi_{\rm in}$ is the initial value at the onset of inflation.

One might think that a natural value for 
$\phi_{\rm in}$ is $m_P$, since we are dealing with a modulus field. However,
such a value would be possible only if the interaction term in the potential
in Eq.~(\ref{V}) is not larger than $V_{\rm inf}$. Hence, using that 
\mbox{$\Phi_0\sim m_P$} we find
\begin{equation}
\lambda\,\phi_{\rm in}^2\Phi_0^2\lsim V_{\rm inf}\quad\Rightarrow\quad
\phi_{\rm in}\lsim\min\left\{\frac{m_{3/2}}{\sqrt{\lambda}},\;m_P\right\},
\label{fin}
\end{equation}
where we also used Eq.~(\ref{Vinf}).
Unless the initial conditions for $\phi$ are tuned to be very close to the 
origin, we expect the above bound to be saturated. 

Now, from Eq.~(\ref{NPhi}) it is evident that, for \mbox{$N_\Phi>0$}, 
we require
\begin{equation}
\lambda>\left(\frac{m_{3/2}}{m_P}\right)^2\sim 10^{-30}.
\label{lbound}
\end{equation}
Hence, Eq.~(\ref{fin}) suggests that\footnote{It is equally possible for the
initial conditions of the system to be 
\mbox{$(\Phi_0, \phi_{\rm in})=(m_{3/2}/\sqrt{\lambda}, m_P)$} instead of 
\mbox{$(\Phi_0, \phi_{\rm in})=(m_P, m_{3/2}/\sqrt{\lambda})$}. In fact, at 
energy \mbox{$\rho\sim M_S^4$} the system has to be in a potential valley 
around either the $\phi$ of the $\Phi$ axis. Selecting the appropriate valley
may be considered tuning by some. However, note that only the valley around the 
$\Phi$ axis (i.e. the axis with \mbox{$\phi=0$}) leads to inflation. So one may
argue that, after inflation, the probability to be in a part of the Universe
corresponding to the appropriate initial conditions is exponentially~large.}
\begin{equation}
\phi_{\rm in}\sim\frac{m_{3/2}}{\sqrt{\lambda}}. 
\label{finfin}
\end{equation}
Therefore, using Eqs.~(\ref{Nfr1}), (\ref{NPhi}), (\ref{b}), (\ref{f0}) and 
(\ref{finfin}) we obtain:
\begin{equation}
\ln\beta\simeq-\frac{3}{4}N_\Phi-\ln\sqrt{\lambda}
\simeq-\left(1+\frac{3}{4\bar F_\Phi}\right)\ln\sqrt{\lambda}-
\frac{3}{4\bar F_\Phi}\ln\left(\frac{m_P}{m_{3/2}}\right).
\label{lnb}
\end{equation}

\subsubsection{\boldmath The new bound on $m_\phi$}

Incorporating Eq.~(\ref{lnb}) into Eq.~(\ref{newbound}), the bound on 
$m_\phi$ is recast as
%
%
\begin{eqnarray}
\frac{m_\phi}{H_{\rm inf}} & < & \frac{3}{2}
\left\{
\left[
\left(
\frac{108+\ln\sqrt{g}-\frac{3}{2}N_0}{\ln(m_P/m_{3/2})}-\frac{3}{4}
-\frac{3}{2\bar F_\Phi}
\left(
1+\frac{\ln\sqrt\lambda}{\ln(m_P/m_{3/2})}
\right)
\right)^{-1}\times
\right.\right.\nonumber\\
& & \times
\left.\left.
\left(
1+\frac{3}{4\bar F_\Phi}
\right)
\left(
1+\frac{\ln\sqrt\lambda}{\ln(m_P/m_{3/2})}
\right)+1
\right]^2-1
\right\}^{1/2}.
\label{finbound}
\end{eqnarray}

Moreover, inserting Eq.~(\ref{lnb}) into Eq.~(\ref{lnbrange}) we find
\begin{equation}
-\ln\left(\frac{m_P}{m_{3/2}}\right)<\ln\sqrt{\lambda}<
-\left(1+\frac{4}{3}\bar F_\Phi\right)^{-1}\ln\left(\frac{m_P}{m_{3/2}}\right)
\leq-\frac{1}{3}\ln\left(\frac{m_P}{m_{3/2}}\right),
\label{lrange}
\end{equation}
which is consistent with the bound in Eq.~(\ref{lbound}) and for the last 
inequality we used Eq.~(\ref{barF}). Note that the upper bound in the above 
demands \mbox{$\lambda\leq 10^{-10}$}.

The upper bound in Eq.~(\ref{lrange}) results in a lower bound on $m_\Phi$.
It is easy to see that this bound, for a given $\lambda$, reads
\begin{equation}
\frac{m_\Phi}{H_{\rm inf}}>\frac{3}{2}\left[1-\frac{1}{4}
\left(3+\frac{\ln(m_P/m_{3/2})}{\ln\sqrt\lambda}\right)^2\right]^{1/2}.
\label{mPhibound}
\end{equation}
If $m_\Phi$ were smaller than the above bound then $N_\Phi$ would be too 
large, which would allow enough time for $\bar\phi$ to decrease below $m_\phi$ 
by the time of the phase transition, resulting into copious PBH production.

\subsubsection{\boldmath Example case: $\lambda\sim\sqrt\alpha$}

To illustrate the above somewhat clearer let us choose 
\begin{equation}
\lambda\sim\sqrt\alpha\sim\frac{m_{3/2}}{m_P}\sim 10^{-15}
\label{leg}
\end{equation}
We immediately see that, according to Eq.~(\ref{mPhibound}), the bound on 
$m_\Phi$ is
\begin{equation}
\frac{m_\Phi}{H_{\rm inf}}>\frac{3\sqrt 3}{4}\approx 1.3\;,
\label{mPhiboundeg}
\end{equation}
which corresponds to 
\begin{equation}
\frac{3}{4}<\bar F_\Phi\leq\frac{3}{2}
\label{barFeg}
\end{equation}

As far as the bound on $m_\phi$ is concerned, it is 
straightforward to show that Eq.~(\ref{finbound}) becomes
\begin{equation}
\frac{m_\phi}{H_{\rm inf}}<\frac{3}{2}
\left\{\left[\left(
\frac{108+\ln\sqrt{g}-\frac{3}{2}N_0}{\ln(m_P/m_{3/2})}-
\frac{3}{4}\left(1+\frac{1}{\bar F_\Phi}\right)\right)^{-1}
\frac{1}{2}\left(1+\frac{3}{4\bar F_\Phi}\right)
+1\right]^2-1\right\}^{1/2},
\label{mphibound}
\end{equation}
while Eq.~(\ref{lnb}) gives
\begin{equation}
\ln\beta\simeq\frac{1}{2}
\left(1-\frac{3}{4\bar F_\Phi}\right)
\ln\left(\frac{m_P}{m_{3/2}}\right)>0,
\label{lnbeg}
\end{equation}
where we also used Eq.~(\ref{barFeg}).

\begin{figure}
\begin{center}
\leavevmode
\hbox{%
\epsfxsize=5in
\epsffile{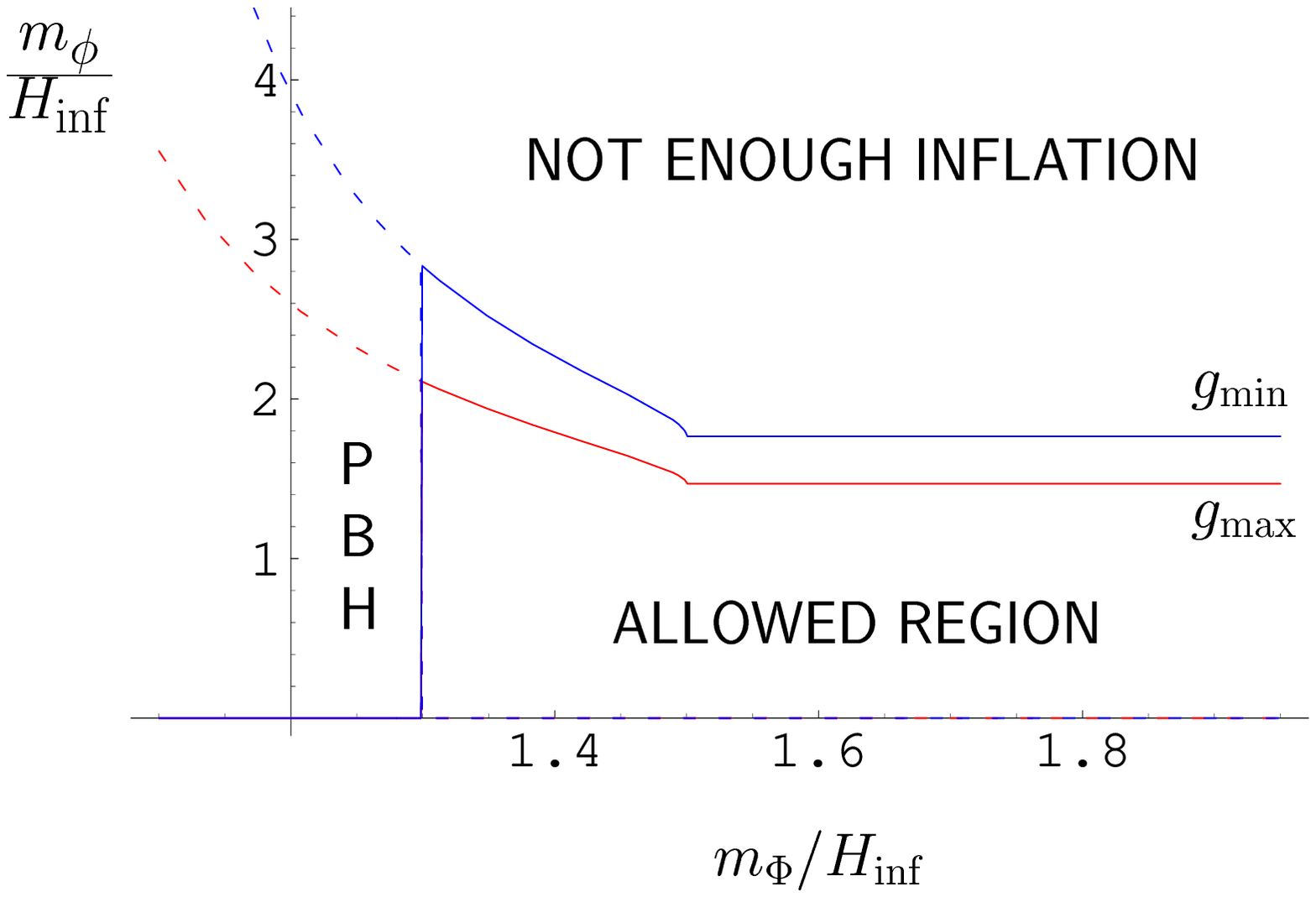}}
\vspace{-9cm}
\caption{
Plot of the allowed parameter space for the masses of the moduli $\Phi$  and 
$\phi$ in order for hybrid inflation without flat directions to work, in the 
case when \mbox{$\lambda\sim\sqrt{\alpha}$} and \mbox{$N_0=0$}. The solid
lines denote the borders of the allowed region. In particular, the upper 
\{lower\} line corresponds to the minimum \{maximum\} value of $g$, i.e.
\mbox{$g_{\rm min}\sim m_{3/2}/m_P$} \{\mbox{$g_{\rm max}\sim 1$}\} according 
to Eq.~(\ref{grange}). The region above the border lines is excluded because
$m_\phi$ is too large to allow an adequately long tachyonic fast-roll inflation
phase . Hence, this region corresponds to inflation with not enough e-foldings 
to explain the cosmological observations (i.e. solve the flatness and horizon 
problems). The region on the left of the the border line is excluded because
$m_\Phi$ is small enough to result in a long first stage of fast-roll 
inflation, which enables $\phi$ to roll too much toward the origin. Hence, in
this region, the tachyonic fluctuations at the end of the first stage of 
inflation dominate the motion of $\phi$ and result in cosmologically 
catastrophic production of Primordial Black Holes (PBHs). The two forbidden 
regions are separated by a dashed line, whose position depends again on the 
value of $g$. The distinction between a first stage of fast roll or locked 
inflation is evident. For \mbox{$m_\Phi>\frac{3}{2}H_{\rm inf}$} we have 
locked inflation, in which case the upper bound on $m_\phi$ is independent of 
$m_\Phi$ [cf. Eq.~(\ref{mphiboundeg})]. This bound is relaxed for smaller 
$m_\Phi$, when we have a first stage of fast roll inflation. Avoiding PBH 
production bounds $m_\Phi$ from below as \mbox{$m_\Phi> 1.3\,H_{\rm inf}$}, 
according to Eq.~(\ref{mPhiboundeg}). We see that the allowed parameter space 
corresponds to \mbox{$m_\phi\sim m_\Phi\sim H_{\rm inf}$} as required.
}
\end{center}
\end{figure}

If we further set \mbox{$m_\Phi>\frac{3}{2}H_{\rm inf}$}, then the first 
stage of inflation 
corresponds to locked inflation and \mbox{$\bar F_\Phi=3/2$}. In this case
Eq.~(\ref{mphibound}) becomes
\begin{equation}
\frac{m_\phi}{H_{\rm inf}}<\frac{3}{2}
\left\{\left[\left(
\frac{72+\frac{2}{3}\ln\sqrt{g}-N_0}{\ln(m_P/m_{3/2})}-\frac{5}{4}\right)^{-1}
\frac{1}{2}+1\right]^2-1\right\}^{1/2},
\label{mphiboundeg}
\end{equation}
which, for \mbox{$N_0=0$} and for the range of $g$ in Eq.~(\ref{grange}), 
interpolates between between 1.5 and 1.8. 
Hence, it is evident that enough inflation to satisfy the 
cosmological observations can be achieved with
\begin{equation}
m_\Phi\sim m_\phi\sim m_{3/2}\sim H_{\rm inf}\;,
\end{equation} 
i.e. without the use of a flat direction. Hence, we have shown that 
{\em the combination of locked and fast-roll inflation is capable of providing 
enough e-foldings of inflation to 
encompass the cosmological scales without the use of any flat direction}. 
Note that the bound in Eq.~(\ref{mphiboundeg}) is further relaxed if 
\mbox{$m_\Phi<\frac{3}{2}H_{\rm inf}$} as can be seen in Figure~3.

Now, regarding the catastrophic possibility of PBH production, 
Eq.~(\ref{lnbeg}), in the case when \mbox{$m_\Phi\geq\frac{3}{2}H_{\rm inf}$}, 
gives
\begin{equation}
\ln\beta\simeq 8.64 \Rightarrow \beta\sim 6\times 10^3,
\end{equation}
which means that $\phi$ is always safely away from the origin so that the 
tachyonic fluctuations never dominate its motion and, consequently, 
{\em there is no excessive production of perturbations and no PBH formation.} 
If \mbox{$m_\Phi<\frac{3}{2}H_{\rm inf}$}, then \mbox{$\bar F_\Phi<3/2$} and 
$\beta$ is reduced, but, as shown in Eq.~(\ref{lnbeg}), it is always greater 
than unity.

In total we have shown that {\em there is ample parameter space in which PBH 
production is avoided while enough inflation is achieved without the use of 
flat directions}. However, there are more requirements to be met for a 
successful cosmology. We discuss the most important of them in the following 
section.

\section{Other cosmological considerations}

\subsection{Curvature perturbations}

There are more to inflation than the solution of the horizon and flatness 
problems. In particular inflation has to provide also the spectrum of 
superhorizon density/curvature perturbations which causes the observed
CMBR anisotropy and seeds the formation of Large Scale Structure in the 
Universe. The superhorizon spectrum of perturbations is thought
to be generated by the amplification of the quantum fluctuations of a light
field, i.e. a field whose mass is smaller than $H_{\rm inf}$. This is because
only if the Compton wavelength of the field is larger than the horizon during 
inflation, can the quantum fluctuations of the field reach and exit the 
horizon, thereby giving rise to the desired generation of a superhorizon 
perturbation spectrum. This is a rather generic requirement, which seems to
call for the use of at least one flat direction in inflation.

Traditionally, it was thought that the field responsible for the generation
of the superhorizon curvature perturbations is the inflaton field. However, 
this is possible only if the inflaton is effectively massless, 
i.e. only if \mbox{$m_\Phi\ll H_{\rm inf}$}. This is in conflict with
our aim in this paper, which is to achieve inflation without the use of flat 
directions, which means that, for our inflaton, 
\mbox{$m_\Phi\sim H_{\rm inf}$}. Consequently an alternative solution must be 
found.

There are a number of recent proposals on this issue in the literature. 
For example, in Ref.~\cite{dvali} it is assumed that $\phi$ is coupled to some 
other field $\chi$ in a way, which does not affect the inflationary scenario 
but does affect the inflaton decay rate $\Gamma$.
The latter is thought to be perturbed on superhorizon
scales because $\chi$ is assumed to be an appropriately light field. This
mechanism was recently introduced in Ref.~\cite{modulG} and, even though quite
interesting, it suffers from the fact that we require our moduli inflatons
to possess the appropriate coupling to the appropriate field, which should
be large enough to generate the required perturbations but not too large
because it should not lift the flatness of the $\chi$ direction. It seems,
therefore, that this mechanism needs a few special requirements to work. This
is against our philosophy, which aims to address the problems of inflation
model--building in the most generic and natural way possible.

Fortunately there is another way to obtain the required curvature 
perturbations. Recently it has been suggested that the generation of
these superhorizon perturbations could be entirely independent of the
inflaton field. Indeed, according to this proposal the curvature 
perturbation spectrum is due to the superhorizon perturbations of some
``curvaton'' field $\sigma$ \cite{curvaton}. As usual, this field must be a 
light field during inflation. Its energy density during inflation is 
negligible and, therefore, has no effect on the inflationary dynamics.
After the end of inflation, however, the field becomes important and
manages to dominate (or nearly dominate) the energy density, 
imposing thereby its own curvature perturbation onto the Universe.
Afterwards, it decays into the thermal bath of the SHBB.

The existence of a curvaton field substantially ameliorates the constraints
imposed by observations onto inflation model-building \cite{liber}. Indeed,
the only requirement that the curvaton imposes onto inflation is that 
the inflaton itself does not generate excessive curvature perturbations. 
In our case, though, there is no such danger because our inflatons are not
light fields and, therefore, they do not produce any sizable superhorizon 
curvature perturbations, because their quantum fluctuations are exponentially
suppressed before reaching the horizon during inflation.

The curvaton is generally expected to produce a quite flat superhorizon
spectrum of curvature perturbations, which is in good agreement with the
recent WMAP observations that suggest \mbox{$n_s\approx 1$} for the spectral 
index. However, the predictions of the curvaton depend on the evolution of the 
field after the end of inflation, which has been thoroughly investigated in
a recent paper of one of us, with collaborators \cite{evol}. In particular, if 
the curvaton decays before it dominates the Universe it may generate a 
substantial isocurvature component on the perturbation spectrum, correlated 
with the adiabatic mode, which may soon become observable by the Planck 
satellite. Also, it is possible to obtain sizable non-Gaussianity that may
be constrained or detected by the SDSS or the 2dF galaxy surveys 
\cite{curvaton}.

The advantage of the curvaton is that, since it is independent of the physics
of inflation, it can be associated with much lower energy scales than 
$V_{\rm inf}^{1/4}$. In particular, it may well be associated with TeV physics,
and can be a field already present in simple extensions of the Standard Model.
Indeed, physics beyond the Standard Model provides a large number of
curvaton candidates such as, a right-handed sneutrino \cite{sneutrino}, used 
for generating the neutrino masses, the Peccei-Quinn field \cite{mine}, which 
solves the strong CP problem, various flat directions of the MSSM 
\cite{mssm}, or of the NMSSM \cite{nmssm}, a PNGB \cite{pngb} (e.g. a Wilson 
line or a string axion) and a string modulus \cite{moroi} to name but some.

\subsection{The moduli problem}

The fact that the SHBB must have already begun before BBN takes place
sets a lower bound on $T_{\rm reh}$ for any inflationary model. This bound 
translates into a bound on $\Gamma$ [cf. Eq.~(\ref{Treh})] which, in turn, 
becomes a bound on $g$ of Eq.~(\ref{G}). Indeed, demanding 
\mbox{$\Gamma\gg H_{\sc bbn}\sim 10^{-24}$ GeV}, we find that BBN constrains
$g$ as follows:
\begin{equation}
g\geq 10\;\frac{m_{3/2}}{m_P}\,.
\label{gcons}
\end{equation}
Thus, we see that almost the entire range in Eq.~(\ref{range}) escapes the 
above bound. There is a problem only if $\phi$ decays through gravitational 
couplings, in which case \mbox{$\Gamma\sim m_{3/2}^3/m_P^2$}. This is the 
well known moduli problem. A typical solution to this problem involves the 
introduction of an additional brief period of thermal inflation 
\cite{thermal}, which dilutes the moduli and produces additional entropy, 
which enables the SHBB to commence earlier than BBN. This period of thermal 
inflation may occur during the
inflaton's coherent oscillations, before reheating is completed. 

Typically, thermal inflation may last up to 20 e-foldings. This means that,
in Eq.~(\ref{Ncosmo}), the term $N_0$ would be non zero, but equal to the
number of e-foldings of thermal inflation. When \mbox{$N_0>0$} the upper bound 
on $m_\phi$ in Eq.~(\ref{finbound}) is relaxed. 
For example, with \mbox{$\lambda\sim\sqrt{\alpha}$}\ and 
\mbox{$m_\Phi>\frac{3}{2}H_{\rm inf}$},
putting \mbox{$N_0=10$} \{\mbox{$N_0=14$}\} relaxes
the bound on the ratio of \mbox{$m_\phi/H_{\rm inf}$} 
[cf. Eq.~(\ref{mphiboundeg})]
up to 5 \{9\}. 
For \mbox{$N_0\rightarrow 17.3$} it can be shown that there is no
need for a second stage of fast-roll inflation and $m_\phi$ can be very 
large as shown in Figure~4.
\begin{figure}
\begin{center}
\leavevmode
\hbox{%
\epsfxsize=5in
\epsffile{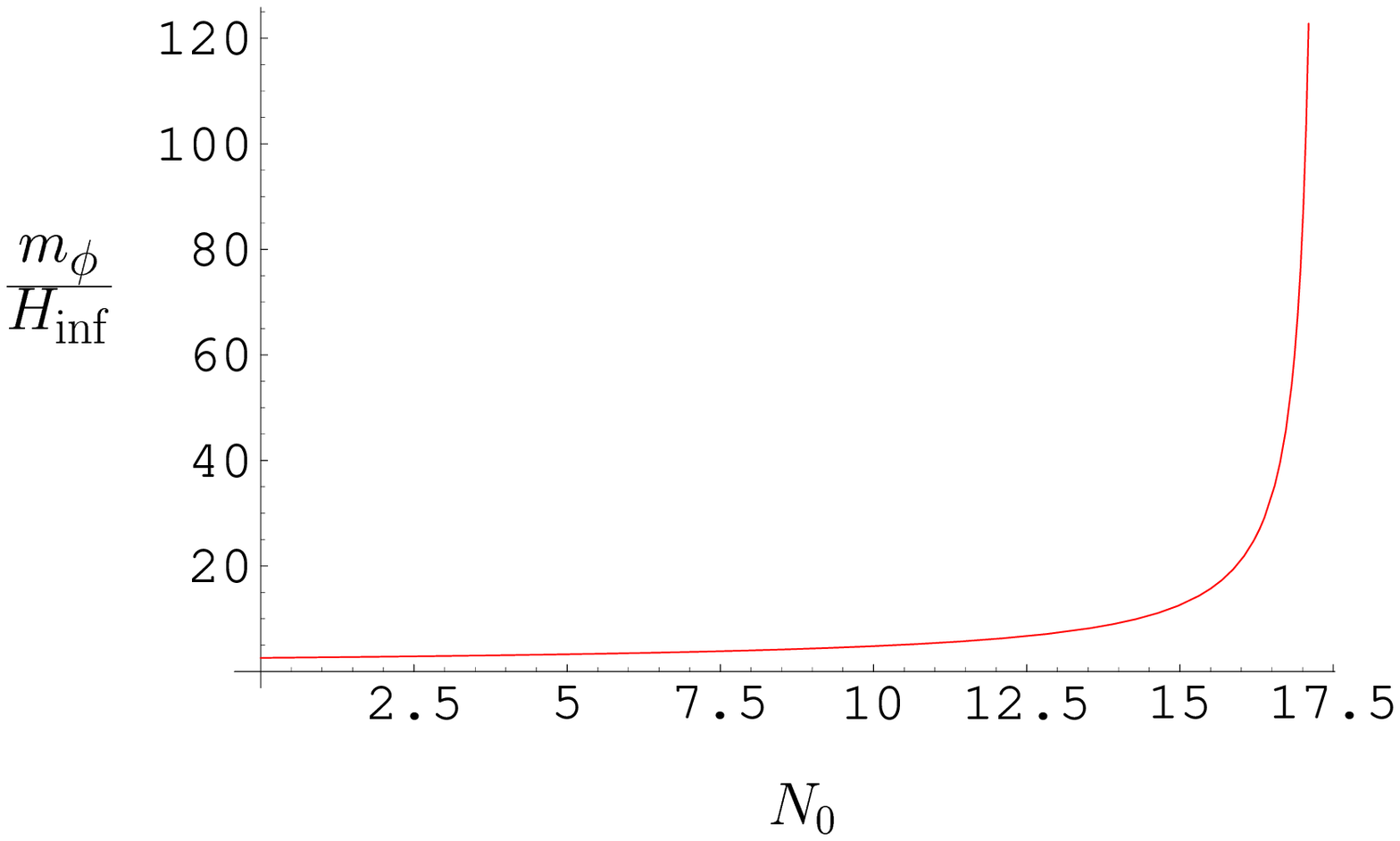}}
\vspace{-9cm}
\caption{
Plot of the upper bound on the ratio 
\mbox{$m_\phi/H_{\rm inf}$} with respect to $N_0$ (the number of e-foldings
corresponding to a brief period of thermal inflation, invoked to overcome the 
moduli problem), in the case when \mbox{$\lambda\sim\sqrt{\alpha}$} and
$\phi$ decays gravitationally, i.e. when 
\mbox{$g\sim m_{3/2}/m_P$}. We see that the larger $N_0$ is the weaker the
bound becomes. When \mbox{$N_0\rightarrow 17.3$} there is no need for any
tachyonic, fast-roll inflation and $m_\phi$ can be arbitrarily large.
}
\end{center}
\end{figure}

\section{Discussion and conclusions}

Using a rather generic form for their scalar potential, we have shown that 
moduli fields, corresponding to flat directions of supersymmetry, whose 
flatness is lifted by supergravity corrections, can naturally generate enough 
e-foldings of inflation to solve the horizon and the flatness problems of the 
Standard Hot Big Bang (SHBB). Indeed, using natural values for the parameters 
(masses of order TeV and vacuum energy of the order of the intermediate scale, 
corresponding to gravity mediated supersymmetry breaking) and a hybrid-type 
potential we have found that the moduli give rise to two-stage inflation whose 
total duration may well be long enough to encompass the cosmological scales. 
Depending on the curvature of the potential the first stage of inflation
may be a period of fast-roll inflation or of oscillatory inflation,
when the system is `locked' on top of an unstable saddle point corresponding
to non-zero vacuum density. This is followed by a second stage of tachyonic
fast-roll inflation, when the system rolls toward the true vacuum. Our 
calculations have demonstrated that, with a quite mild upper bound on the 
tachyonic mass of the inflaton, we can achieve enough e-foldings of inflation 
without employing slow roll at all. That way inflation can escape from the 
famous $\eta$-problem, since we manage to have masses of order the Hubble 
parameter without problem.

Probably the most difficult obstacle to the success of our scenario is the 
possibility of copious generation of of Primordial Black Holes (PBHs). They may be generated due to excessive tachyonic fluctuations at the phase transition,
which terminates the first stage of inflation. We have studied this problem in 
detail and showed that, if the PBHs do form then it is impossible to return to 
the SHBB cosmology in time for BBN. The only solution is, therefore, 
to avoid creating the PBHs in the first place. By considering the initial 
conditions of $\phi$ more carefully and by following its evolution during the 
first stage of inflation we have demonstrated that it is indeed possible to 
prevent it from being, at the time of the phase transition, under the influence
of excessive tachyonic fluctuations, which would lead to PBH production. 
Instead, we have shown that, with natural initial conditions, one can avoid the
PBHs provided $\lambda$ is not very large 
(\mbox{$\lambda\sim 10^{-10}$} at most), i.e. the interaction between the moduli
is suppressed. This is quite likely for moduli fields away from enhanced 
symmetry points (especially if the coupling is controlled by the 
Planck--suppressed VEV of some other field). Note, also, that avoiding the 
PBHs in the way we propose also dispenses with another potential danger; that 
of generating cosmologically catastrophic topological defects at the phase 
transition. For example, were it otherwise, it would be possible to generate 
stable domain walls, which would disastrously dominate the Universe. Still, 
one can consider more complicated theories where such defects are unstable or 
harmless (e.g. cosmic strings at the energy scale $M_S$ have little 
cosmological consequences).

Structure formation, in our model, is due to the existence of a curvaton field,
which is unrelated to the moduli inflatons and, consequently, it can neither
affect the dynamics of inflation nor does it have to be tuned accordingly to 
avoid this danger. The curvaton {\em must} be a flat direction because there 
is no known way to obtain the superhorizon spectrum of density perturbations 
required by the observations, other than inflating the vacuum fluctuations of
a light scalar field. However, since the curvaton is not related with the 
inflationary expansion, it can be protected by some symmetry, which may even be
exact during inflation (e.g. a global U(1) for a PNGB curvaton). Moreover, 
the curvaton can be associated with low energy (TeV) physics and can be easily
accommodated in simple extensions of the Standard Model. 

As discussed also in Ref.~\cite{dvali} the potential landscape for the moduli 
fields is expected to allow a cascade of periods of oscillatory, `locked' 
inflation. Completing this picture we add that, between those periods, we 
can easily have periods of fast-roll inflation when the system is rolling from 
one saddle point to another. That way the total number of e-foldings can be
much larger than the one corresponding to the cosmological scales. Of course 
one needs a roughly flat region of the Universe to start up with, but this is a
generic initial condition problem for inflation. This work shows that, at 
least, the other generic problem of inflation, namely the $\eta$-problem, can 
be naturally overcome.

\bigskip

\noindent
{\large\bf Acknowledgements:} We would like to thank David H. Lyth for 
discussion and comments.

\end{document}